\pdfoutput=1
\documentclass[pra,aps,superscriptaddress,amsmath,amssymb,nofootinbib]{article}

\usepackage{authblk, comment}

\usepackage{physics}
\usepackage{braket,verbatim,bm,bbold,epsfig,float,color}
\usepackage[a4paper, total={6in, 8in}]{geometry}
\usepackage[colorlinks=true,linkcolor=blueG, citecolor=redG, urlcolor=magentaG, bookmarks]{hyperref}
\usepackage{color}
\usepackage[dvipsnames]{xcolor}
\usepackage{enumitem}  
\usepackage{amsmath, amssymb, amsfonts, mathtools}
\usepackage{graphicx}
\usepackage{caption}
\usepackage{subcaption} 
\usepackage{amsfonts} 
\usepackage{cite}
\usepackage{tensor}
\newcommand{\mytensor}[1]{\overset{\scriptscriptstyle\leftrightarrow}{#1}}

\definecolor{blueG}{RGB}{51, 102, 204}
\definecolor{magentaG}{RGB}{214.2, 40.8, 132.6}
\definecolor{redG}{RGB}{229.5, 51., 102.}

\newcommand{\nc}{\newcommand}
\nc{\Hc}{\mathcal{H}}
\nc{\Hent}{\mathcal{H}_{ent}}
\nc{\A}{\mathcal{A}}
\nc{\x}{\chi}
\newcommand{\ir}{\mathrm{i}}
\newcommand{\eE}{\mathrm{e}}

\newcommand{\appropto}{\mathrel{\vcenter{
  \offinterlineskip\halign{\hfil$##$\cr
    \propto\cr\noalign{\kern2pt}\sim\cr\noalign{\kern-2pt}}}}}

\begin{document}

%%%%%%%%%%%%%%%%%%%%%%%%%%%%%%%%%%%%%%%%%%%%%%%%%%%%%%%%%%%%%%%%%%%
\title{\bf Entanglement Hamiltonian and orthogonal~polynomials}
%%%%%%%%%%%%%%%%%%%%%%%%%%%%%%%%%%%%%%%%%%%%%%%%%%%%%%%%%%%%%%%%%%%
\author[1]{Pierre-Antoine Bernard}
\author[2]{Riccarda Bonsignori}
\author[2]{Viktor Eisler}
\author[3]{Gilles Parez}
\author[2,4]{Luc Vinet}

\affil[1]{\it Centre de Recherches Math\'ematiques (CRM), Universit\'e de Montr\'eal, P.O. Box 6128, Centre-ville
Station, Montr\'eal (Qu\'ebec), H3C 3J7, Canada,}

\affil[2]{\it  Institute of Theoretical and Computational Physics, Graz University of Technology,
Petersgasse 16, 8010 Graz, Austria,}

\affil[3]{\it Laboratoire d’Annecy-le-Vieux de Physique Théorique (LAPTh), CNRS, Université Savoie Mont Blanc, 74941 Annecy, France,}

\affil[4]{\it IVADO,  Montréal (Québec), H2S 3H1, Canada.}

\date{\today}

\maketitle

\begin{abstract}
We study the entanglement Hamiltonian for free-fermion chains with a particular form of inhomogeneity. The hopping amplitudes and chemical potentials are chosen such that the single-particle eigenstates are related to discrete orthogonal polynomials of the Askey scheme. Due to the bispectral properties of these functions, one can construct an operator which commutes exactly with the entanglement Hamiltonian and corresponds to a linear or parabolic deformation of the physical one. We show that this deformation is interpreted as a local inverse temperature and can be obtained in the continuum limit via methods of conformal field theory. Using this prediction, the properly rescaled eigenvalues of the commuting operator are found to provide a very good approximation of the entanglement spectrum and entropy.
\end{abstract}

\textbf{Keywords}: Quantum entanglement, orthogonal polynomials, conformal field theory, free fermions
\tableofcontents

\section{Introduction}\label{sec:intro}

Entanglement is a fundamental phenomenon in quantum mechanics, and its description and quantification have become central topics of investigation across various fields of physics \cite{AFOV-08,Laf-16,Wit-18, Nis-18}. In the context of many-body quantum systems, a key object is the reduced density matrix (RDM). For bipartite pure states, it fully encodes the entanglement between a given subsystem and its complement, and is obtained by tracing out the degrees of freedom of the latter from the density matrix of the whole system. In order to characterize the correlations between the subsystems, one can adopt the system-bath formalism of statistical mechanics, and rewrite the RDM as $\rho_{\mathcal{A}}=\exp(-\mathcal{H})/Z$. The operator $\mathcal{H}$, known as entanglement Hamiltonian (EH), encodes all the information on the RDM in a more suitable form, and its theoretical characterization has been a topic of intense studies \cite{DEFV-23}. 
This has been accompanied, in recent years, by the development of experimental protocols that allowed the reconstruction of the EH in quantum simulator experiments \cite{Kokailetal21a, Joshietal23}. All these remarkable results have been achieved thanks to the observation that, in many relevant cases, the EH can be written as a local operator. In particular, in the ground state of a broad class of many-body systems, it has the form of a spatial deformation of the physical Hamiltonian.

In the framework of relativistic quantum field theories (QFT), this property is grounded in the Bisognano-Wichmann (BW) theorem \cite{BW75,BW76}. According to BW,  the EH (or modular Hamiltonian) of a semi-infinite system is given by the generator of Lorentz boosts in the direction perpendicular to the boundary of the half-space, and thus corresponds to a linear deformation of the Hamiltonian.
In the presence of additional conformal symmetry, this result has been generalised to further geometries~\cite{HL-82, CHM-11}. In particular, for $(1+1)$ dimensional conformal field theories (CFT), the EH can be written under certain conditions as a local integral over the energy-density component of the stress tensor,
\begin{equation}
\label{eq:BWth}
\mathcal{H}= 2 \pi \int_\A \dd x \beta(x) T_{00}(x),
\end{equation}
where the weight function $\beta(x)$ depends on the geometry of the subsystem $\A$ \cite{CT-16}. The notation is chosen to make an explicit reference  to the inverse temperature in the thermodynamic analogy. 

The expression of the EH as spatial deformation of the physical Hamiltonian, remarkably, holds in many examples beyond relativistic field theories. A notable example is the one of non-critical integrable chains, where the EH of a semi-infinite system is identified with the corner transfer matrix of a corresponding two-dimensional statistical physics model \cite{PKL-99}. In this geometry, the EH is exactly given by the discretized version of \eqref{eq:BWth} \cite{Eis-24}. Moreover, this ansatz is found to give an excellent approximation for more general lattice models, whose low-energy behavior is described by relativistic QFTs \cite{GMSCD-18,MSGDR-18, DVZ-18}.
For the explicit study of the lattice EH, free-fermion systems provide a particularly suitable framework \cite{PE-09}.
For the homogeneous hopping chain, the analytical expression of the lattice EH has been derived \cite{EP-17,EP-18}, and shows deviations from the CFT predictions due to the presence of long-range hopping terms.
However, as subsequently shown, taking a proper continuum limit of the lattice EH allows to recover the field theory prediction \cite{ABCH-17,ETP-19}.
These hopping chains show a further interesting property, namely their EH is found to commute with a tridiagonal matrix that has exactly the discretized form of Eq.~\eqref{eq:BWth}\cite{Pes-04,EP-13}. Analogous properties have also been found in the case of non-relativistic continuum models such as the free Fermi gas, where the commuting operator is a second-order differential operator \cite{Eis-23,Eis-24}. The existence of such commuting operators is an interesting feature, since it reveals connections with apparently distant fields as the classical theory of signal processing \cite{BCNPV-24}, enabling the application of techniques and results originally developed in these contexts\cite{Slepian-78,SP-61,Landau-85}.

The criteria underlying the existence of such commuting operators, while not yet fully understood, have been systematically explored in a series of works \cite{Grunbaum-81,DG-86,Gru-01}. These suggest that  a key role is played by the  bispectrality property of the eigenfunctions of the total system. 
Subsequent works by Grünbaum, Vinet and Zhedanov showed that in each bispectral problem it is possible to construct the so-called algebraic Heun operator, which is the commuting operator for the time- and band-limiting problem \cite{GVZ-17, GVZ-18}. This has been employed in the study of inhomogeneous hopping chains associated to discrete orthogonal polynomials to construct a matrix commuting with the EH \cite{crampe2019free, crampe2021entanglement}. However, the relation between the corresponding spectra was left unexplored.

Further examples of inhomogeneous free-fermion systems with a commuting operator were studied recently in \cite{BE-24}. In particular, the commuting operator was understood as a deformation of the physical Hamiltonian, in terms of an effective inverse temperature that follows from a CFT treatment in the continuum limit \cite{TRLS-18}. Moreover, this relation even provides the missing link between the spectra of the EH and the commuting operator, such that their eigenvalues are related, to a good approximation, via a proper rescaling. This motivates further investigation of such a correspondence for hopping chains with an underlying bispectrality property \cite{crampe2021entanglement}. %To this purpose, orthogonal polynomials appear as ideal candidates, because of their bispectral properties.

In the present paper, 
we extend this framework to hopping chains with inhomogeneous couplings and potentials, associated to families of orthogonal polynomials of the so-called Askey scheme, a hierarchical classification of the hypergeometric polynomials \cite{koekoek2010hypergeometric}. This scheme identifies~in~particular all discrete orthogonal polynomials whose bispectral properties involve satisfying a three-term difference relation in addition to a three-term recurrence relation \cite{terwilliger2004leonard}.
Beyond their bispectral properties, these chains are interesting for their physical applications, notably in the context of perfect state transfer. This protocol, which aims at the end-to-end transfer of a quantum state through the dynamics of a spin chain, can be implemented using the Krawtchouk chain \cite{christandl2004perfect}. In the following, we consider the Krawtchouk, dual Hahn, Hahn and Racah families of 
orthogonal polynomials~\cite{crampe2019free,BCNPdAV-22,VZ-19,Blanchet_2024,bernard2024distinctive}. First, we consider the continuum limit of the chains and apply the techniques of \cite{TRLS-18} to obtain the CFT prediction for the inverse temperature for non-symmetric Fermi velocities. Second, for each family, we show that the commuting operator has the expected form of a linear or parabolic deformation of the physical Hamiltonian. Moreover, with the proper rescaling in terms of the Fermi velocity at the boundary, we show that the relevant spectrum of the EH can be well reproduced. 

The paper is organized as follows. In Sec.~\ref{sec:model} we introduce the free fermionic lattice Hamiltonians associated to discrete orthogonal polynomials of the Askey scheme, and illustrate the construction of the associated Heun operators. We also formulate the continuum limit for these models. In Sec.~\ref{sec:CFT} we present the CFT derivation of the EH, specializing to the case of free Dirac fermion theories in curved space. In Sec. \ref{sec:examples}, we proceed to present the results for the lattice models considered, associated to various families of  orthogonal polynomials. In Sec.~\ref{sec:out}, we discuss the results and present some future outlooks. Finally, in the Appendices~\ref{sec:appA} and \ref{app:racah}, we report some details of the calculations.

\section{Model and methods}
\label{sec:model}
In this section, we introduce the free fermionic lattice models of interest, and their relation to orthogonal polynomials. We then define the ground-state EH and, using the bispectrality of the hypergeometric orthogonal polynomials, we construct an operator that commutes with it. Finally, we discuss the continuum limit of the inhomogeneous hopping chain.

\subsection{Hopping chains and orthogonal polynomials}
We consider spinless fermions hopping on an open chain with inhomogeneous couplings $J_n$ and potentials $B_n$. The Hamiltonian is
\begin{equation}\label{def:Hamil}
     \widehat{H} = \sum_{n=0}^{N-2}J_n (c_n^\dagger c_{n+1}
     + c_{n+1}^\dagger c_n) - \sum_{n=0}^{N-1} B_n c_n^\dagger c_n,
\end{equation}
where $c_n^{(\dagger)}$ are fermionic creation and annihilation operators, $N$ is the length of the chain and by convention we start labeling the sites at $n=0$.
Because the Hamiltonian \eqref{def:Hamil} is quadratic in the fermionic operators, it can be diagonalized in terms of new fermionic operators $b_k^\dagger$ and $b_k$ using the canonical transformation
\begin{equation}
     b_k = \sum_{n=0}^{N-1} \phi_n(\epsilon_k) c_n, \qquad b_k^\dagger = \sum_{n=0}^{N-1}  \phi_n(\epsilon_k) c_n^\dagger.
\end{equation}
The single-particle wavefunctions $\phi_n(\epsilon_k)$ are given by the eigenvectors of the hopping matrix, and they satisfy the following three-term recurrence relation
\begin{equation}\label{eq:rec_phi}
    \epsilon_k \phi_n(\epsilon_k) = J_{n} \phi_{n+1}(\epsilon_k) - B_n \phi_n(\epsilon_k) + J_{n-1}\phi_{n-1}(\epsilon_k).
\end{equation}
Here, $\epsilon_k$ is the single-particle spectrum, and the Hamiltonian is recast as
\begin{equation}\label{eq:diag_hamil}
    \widehat{H} = \sum_{k=0}^{N-1} \epsilon_k b_k^\dagger b_k.
\end{equation}

Our goal is to consider a particular form of the couplings $J_n$ and potentials $B_n$ that give rise to eigenvectors $\phi_n(\epsilon_k)$ with bispectral properties. Following \cite{crampe2019free}, we focus on cases where they are related to
discrete orthogonal polynomials of the Askey scheme \cite{koekoek2010hypergeometric}. We denote by $R_n(\epsilon_k)$ a family of such polynomials of degree $n$ in the variable $\epsilon_k$, with $n,k=0,1,\dots,N-1$.
They are orthogonal with respect to the weights $W_k>0$, have norm squared $U_n>0$, 
\begin{equation}\label{eq:ortho}
    \sum_{k=0}^{N-1} W_kR_m(\epsilon_k)R_n(\epsilon_k) = U_n \delta_{mn},
\end{equation}
and satisfy a recurrence relation of the form 
\begin{equation}\label{eq:rec}
    \epsilon_k R_n(\epsilon_k) = A_n R_{n+1}(\epsilon_k)-(A_n+C_n)R_n(\epsilon_k)+C_n R_{n-1}(\epsilon_k),
\end{equation}
where $A_n,C_n$, 
are real parameters satisfying $C_0 = A_{N-1}=0$. Moreover, they have the 
property of satisfying an additional three-term difference relation,
\begin{equation}\label{eq:bi2}
%\begin{split}
     \x_n R_n(\epsilon_k)  = \bar{A}_k R_{n}(\epsilon_{k+1}) - \left(\bar{A}_k  + \bar{C}_k\right)R_{n}(\epsilon_k)
     + \bar{C}_k  R_{n}(\epsilon_{k-1}),
%\end{split}
\end{equation}
with  $\bar{C}_0 = \bar{A}_{N-1}=0$. The coefficients $A_n$, $C_n$, $\bar{A}_k$ and $\bar{C}_k$, as well as the eigenvalues $\varepsilon_k$ and $\chi_n$, can be found in \cite{koekoek2010hypergeometric} for the various families of discrete orthogonal polynomials of the Askey scheme. 

In order to relate the recurrence relation \eqref{eq:rec} to the eigenvalue equation \eqref{eq:rec_phi}, one needs an additional symmetrization. The idea is to renormalize the polynomials $R_n(\epsilon_k) \to \phi_n(\epsilon_k) $ in such a way that the coefficients of $\phi_{n+1}$ and $\phi_{n-1}$ in the new recurrence relation are obtained from a single function $J_n$, evaluated in $n$ and $n-1$, respectively. As shown in \cite{crampe2019free}, the couplings and potentials of the hopping chain are given by
\begin{equation}\label{def:JnBn}
    J_n = \sqrt{A_n C_{n+1}}, \quad B_n = A_n + C_n,
\end{equation}
whereas the eigenvectors are related as
\begin{equation}
\phi_n(\epsilon_k)  = \sqrt{W_k}\sqrt{\frac{A_{n-1} \dots A_0}{C_n \dots C_1}} R_n(\epsilon_k).
    \label{phiR}
\end{equation}
Hence, the solutions of the recurrence relations \eqref{eq:rec_phi} and \eqref{eq:rec} are related by a similarity transformation that preserves the spectrum $\epsilon_k$. Note, however, that the orthonormality of the wavevectors $\phi_n(\epsilon_k)$ requires to include also the weight factor $\sqrt{W_k}$ into their definition.

We have thus demonstrated that, choosing the inhomogeneity pattern according to \eqref{def:JnBn}, the Hamiltonian \eqref{def:Hamil} is solved using discrete orthogonal polynomials of the Askey scheme. Note that this construction includes the homogeneous chain, associated with the discretization of Chebyshev polynomials. Its derivation from the case associated to $q$-Racah polynomials is a nontrivial process involving a truncation and specialization of the Askey-Wilson polynomials at $q$ a root of unity, as discussed in \cite{spiridonov1997zeros}. 
In the next step, we consider the EH for these models and recall the role of the bispectrality of the polynomials $R_n(\epsilon_k)$ in constructing a commuting operator.

\subsection{Entanglement Hamiltonian and commuting operator}
In the ground state of a fermionic hopping chain, the reduced density matrix $\rho_{\A}=e^{-\mathcal{H}}/\mathcal{Z}$ of subsystem $\A$ is described by the EH~\cite{peschel2003calculation},
\begin{equation}
\label{eq:EHop}
    \mathcal{H} = \sum_{i,j \in \A} h_{ij} c_i^\dagger c_j, 
\end{equation}
which is a free-fermion operator characterized by the entries $h_{ij}$ of a hopping matrix
\begin{equation}
\label{eq:EH}
     h = \ln \left(\frac{1-C_\A}{C_\A}\right),
\end{equation}
that follows from the truncated correlation matrix $C_\A$, with entries $i,j \in \A$ given by
\begin{equation}
     C_{ij} = \sum_{\epsilon_k < 0}%\substack{ k\ s.t. \\ \epsilon_k < 0}}
     \phi_i(\epsilon_k) \phi_j(\epsilon_k).
\end{equation}

The truncated correlation matrix $C_\A$ for chains associated to polynomials of the Askey-scheme admits a simple tridiagonal commuting matrix $T_\A$ \cite{crampe2019free,crampe2021entanglement}. This is due to the bispectrality of the polynomials $R_n(\epsilon_k)$, which satisfy both three-term recurrence \eqref{eq:rec} and difference \eqref{eq:bi2} relations. To construct the commuting operator, 
let us first define the single-particle Hamiltonian. In the position basis, i.e., the basis of localized single-particle states, it takes the following tridiagonal matrix form:
\begin{equation}\label{eq:matH}
\begin{split}
    H =  \begin{pmatrix}
        -B_0 & J_0 &  &  & &   \\
        J_0 & -B_1 & J_1 &  &  &   \\
      & J_1 & -B_2 & J_2 &  &   \\
         &  & \ddots & \ddots & \ddots &  \\
        &  &   &   & -B_{N-2} & J_{N-2}\\
       &  &  &  &   J_{N-2} & -B_{N-1}   \\
    \end{pmatrix}.
    \end{split}
\end{equation}
Let us also define the following position operator
\begin{equation}\label{eq:matX}
    X = \text{diag}(\x_0, \x_1, \dots \x_{N-1} ),
\end{equation}
which is diagonal in the position basis, with eigenvalues given by the dual spectrum associated with the difference relation \eqref{eq:bi2}. To proceed, we introduce an operator $T$ which is tridiagonal in the eigenbases of both $H$ and $X$, such that its restriction to $\A$ commutes with the chopped correlation matrix.
For a ground state composed of filled levels $k \in \{0,1,\dots, K\}$ and a subsystem chosen as the segment ending at the boundary, 
$\A = \{\ell+1, \cdots, N-1\}$, this operator is \cite{crampe2019free}
\begin{equation}\label{def:T}
       T = \frac{1}{2}\{H - \frac{\epsilon_{K+1} + \epsilon_{K}}{2},X - \frac{\x_{\ell+1} + \x_{\ell}}{2}\},
\end{equation}
where $\{O_1,O_2\}=O_1O_2+O_2O_1$ is the anticommutator of $O_1$ and $O_2$.
The shift of the energy operator $H$ in the definition \eqref{def:T} can be interpreted as a chemical potential
\begin{equation}
    \mu_0 = \frac{\epsilon_K + \epsilon_{K+1}}{2},
\end{equation}
that lies halfway between the last occupied and first empty level. 

The $T$ matrix has a tridiagonal structure
\begin{equation}
T=\begin{pmatrix}
\label{eq:Ttrid}
d_0 & t_0 & & & & \\
t_0 & d_1 & t_1 & & &\\
 & t_1 & d_2 & t_2 & &\\
  & & \ddots & \ddots& \ddots &\\ 
  & & & &d_{N-2} &t_{N-2} \\
   & & & &t_{N-2} &d_{N-1} 
\end{pmatrix},
\end{equation}
and using Eqs.~\eqref{eq:matH} and \eqref{eq:matX}, its entries in the position basis are obtained as
\begin{equation}
\label{eq:Telements}
    \begin{split}
        d_{i} &=   - \left(  \x_i - \frac{\x_{\ell} + \x_{\ell+1}}{2}\right) (B_i +  \mu_0) ,\\ 
    t_{i} &= \left(\frac{\x_i + \x_{i+1}}{2} - \frac{\x_\ell + \x_{\ell+1}}{2}\right)J_i.
    \end{split}
\end{equation}
From $t_\ell=0$ one can see immediately that $T$ decouples into two matrices supported on $\A$ and its complement $\bar \A$. This follows from the shift of the position operator $X$ in  the definition of $T$, analogous to the shift induced by the chemical potential that sets the Fermi sea. Using the bispectral relations \eqref{eq:rec} and \eqref{eq:bi2}, one can then show that the matrices $T_\A$ and $T_{\bar\A}$ commute with the respective truncated correlation matrices, $[T_\A,C_\A]=[T_{\bar\A},C_{\bar\A}]=0$ \cite{crampe2019free}. Note that the matrix elements \eqref{eq:Telements} differ from those found in \cite{crampe2019free} by an additive constant and the factor $1/2$ in the definition \eqref{def:T}.

The commutation property together with \eqref{eq:EH} ensures, that these matrices share a common eigenbasis,
\begin{equation}
\label{eq:spectra}
C_\A \ket{\psi_k} = \zeta_k \ket{\psi_k}, \qquad
T_\A \ket{\psi_k} = \lambda_k \ket{\psi_k}, \qquad
h \ket{\psi_k} = \varepsilon_k \ket{\psi_k}.
\end{equation}
This observation is useful for the diagonalization of
$C_\A$, which is ill-conditioned and exhibits an accumulation of eigenvalues $\zeta_k$ exponentially close to $0$ and $1$.
In contrast, $T_\A$ typically has a well-spaced spectrum $\lambda_k$, making it more suitable to calculate the common eigenvectors $\ket{\psi_k}$ numerically. Moreover, the quadratic operator
\begin{equation}\label{def:THamil}
    \mathcal{T} = \sum_{i,j \in A} T_{ij} c_i^\dagger c_j
\end{equation}
defines a local Hamiltonian, where the couplings and potentials \eqref{eq:Telements} are given by a deformation of the respective physical parameters.
However, the commutation property alone does not give any information on the relation between $\lambda_k$ and the single-particle entanglement spectrum $\varepsilon_k$.
In order to find a relation between $\mathcal{T}$ and $\mathcal{H}$, we shall consider the continuum limit of the chain and introduce a CFT technique to calculate the EH in the following sections.
 
\subsection{Continuum limit}\label{sec:LDA}

The Hamiltonians \eqref{def:Hamil} associated to discrete orthogonal polynomials of the Askey scheme are described by slowly varying parameters $J_n$ and $B_n$, admitting a smooth limit for large $N$. For the purpose of a later CFT treatment, it will be useful to discuss their continuum limit, which is mainly based on a local density approximation (LDA).
This amounts to introducing a continuously varying coupling $J(x)$ and chemical potential $\mu(x)$, and assuming that the ground state has an effective local description. Namely, instead of considering the global spectrum $\epsilon_k$, we will introduce an effective local dispersion relation, which takes the form of the corresponding homogeneous problem,
\begin{equation}
\label{eq:dispx}
\omega_q(x)=2 J(x) \cos q- \mu(x).
\end{equation}
The local ground state is then characterized by a Fermi sea, with space-dependent Fermi momenta
\begin{equation}
\label{eq:qF}
q_F(x)=\pm\arccos(\mu(x)/2J(x)),  
\end{equation}
\begin{equation}
\label{eq:vF}
v_F(x)=\left|\frac{\mathrm{d}\omega_q(x)}{\mathrm{d}q}\right|_{q_F(x)}=\sqrt{4J^2(x)-\mu^2(x)}.
\end{equation}
The Fermi velocity is defined to be positive, as we would like to associate it with right-moving excitations, whereas the opposing Fermi point will describe left-moving ones. Note that, for $v_F(x)$ to be well-defined, one needs to restrict to the domain with $|\frac{\mu(x)}{2J(x)}|<1$. Indeed, if this condition is not satisfied, then the dispersion \eqref{eq:dispx} has no roots, and the ground state is either completely filled or empty. On the level of LDA, these regions are in a trivial product state and thus do not contribute to the entanglement. For a consistent definition, one should set $v_F(x)=0$ on these domains.

The continuum limit can also be introduced by a formal procedure, inserting the lattice spacing $a$ explicitly into the formulas. In particular, the position is measured by $x=(n+1)a$, while the continuous hopping amplitude and the chemical potential in the LDA must be defined as
\begin{equation}
\label{eq:Jmux}
    J_n \rightarrow a^{-1}J(x), \qquad B_n+\mu_0 \rightarrow a^{-1}\mu(x). 
\end{equation}
Note that the factor $a^{-1}$ formally carries the dimension of energy, while the functions $J(x)$ and $\mu(x)$ are dimensionless. Furthermore, for the fermion operators one has to substitute
\begin{equation}
    c_n \rightarrow  \sqrt{a} (e^{i \varphi(x)}\psi_R(x) + e^{-i \varphi(x)}\psi_L(x)),
\end{equation}
where the phase is given by
\begin{equation}
    \varphi(x)=\int^{x}_0 q_F(x') \, \dd x' .
\end{equation}
This choice absorbs the quickly oscillating contributions due to the nonzero Fermi momentum $q_F(x)$ generated by the lattice dispersion \eqref{eq:dispx}, and ensures that $\psi_R(x)$ and $\psi_L(x)$ are slowly varying right- and left-moving fields.
Carrying out the continuum limit $a \to 0$ then amounts to inserting these substitutions into \eqref{def:Hamil}, Taylor expanding, and keeping only the lowest nonvanishing order in the lattice spacing. For simplicity, we shall require that $Na=1$, such that the total chain is mapped onto the unit interval $[0,1]$ in the continuum. Using the LDA expressions \eqref{eq:qF} and \eqref{eq:vF}, it is a standard exercise to show that the continuum limit yields
\begin{equation}
\label{eq:inhomDirac}
    \widehat{H} \xrightarrow[a\rightarrow 0]{}  \int_{0}^1 v_F(x) \, T_{00}(x) \, \dd x,
\end{equation}
where $T_{00}(x)$ is the energy density operator associated to free massless Dirac fermions,
\begin{equation}
    T_{00}(x) = \frac{1}{2}\left[ \psi_R^\dagger(x) (-i\partial_x)\psi_R(x) - \psi_L^\dagger(x) (-i\partial_x)\psi_L(x) + \text{h.c.}\right].
\end{equation}
The continuum limit of the inhomogeneous free-fermion chain \eqref{def:Hamil} is thus given by a massless Dirac theory with a space-dependent Fermi velocity, which is a relativistic model with a nontrivial local metric~\cite{DSVC-17}. 

Before we continue with the presentation of a CFT approach for the calculation of the EH, some comments are in order. First, as we have remarked earlier, the Fermi velocity is, in general, nonvanishing only on a subregion of the total system. In fact, we shall assume that the expression~\eqref{eq:vF} has only two zeros $v_F(x_\pm)=0$
with $x_\pm \in [0,1]$, and thus the theory \eqref{eq:inhomDirac} is effectively defined on the reduced spatial domain $[x_-,x_+]$.
Second, the substitutions in \eqref{eq:Jmux} together with the choice $Na=1$ make it clear that the lattice couplings $J_n$ and $B_n$ should scale with $N$ for a proper continuum limit. 
 Third, it is worth noting that the structure of the \emph{global} spectrum enters only via the constant chemical potential in \eqref{eq:Jmux}, which must be set $\mu_0 = (\epsilon_{K}+\epsilon_{K+1})/2$ to obtain the ground state with $K+1$ particles.

\section{CFT approach}
\label{sec:CFT}

\begin{figure}[t]
\centering 
\includegraphics[width= \textwidth]{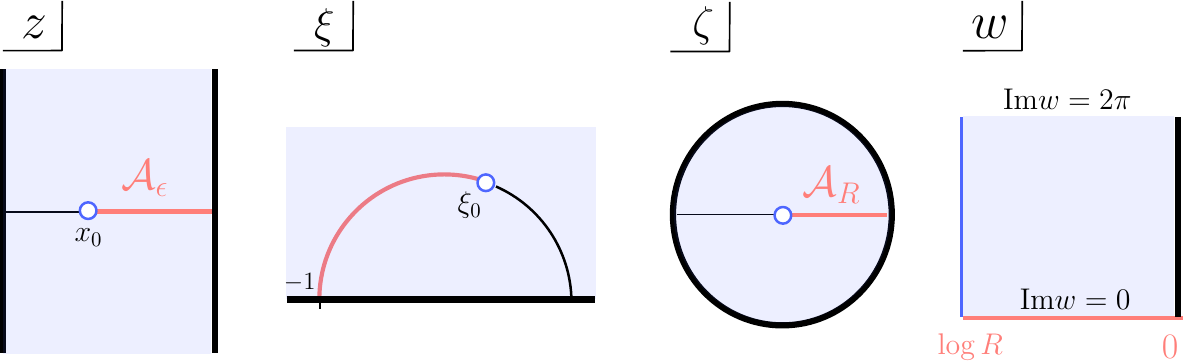} 
     \caption{Conformal mapping to the annulus.}
     \label{fig:map}
\end{figure}

In this section, we present the CFT derivation of the EH for inhomogeneous systems. In particular, we recapitulate the method of \cite{TRLS-18} based on the curved-space CFT formulation~\cite{DSVC-17}, and adapt it to the case where the background metric is not symmetric with respect to the middle of the system.
We start treating the homogeneous case, considering a finite system $x \in [0, L]$ bipartitioned into a subsystem $\A \in [x_0, L]$ and its complement $\bar\A$. 
The RDM is then represented by a path integral on an infinite strip $z\in [0,L]\times \mathbb{R}$ in Euclidean spacetime, with a branch cut on the real axis running along the subsystem $\A$. The corresponding path integral can be regularized by removing an infinitesimal disk $\{|z-x_0|\leqslant \epsilon\}$ around the entangling point, as shown in the first panel of Fig.\ref{fig:map}. With this procedure, we ensure that the entanglement entropy of the regularised interval $\A_{\epsilon}=[x_0+\epsilon,L]$ is finite.

In order to derive the CFT expression of the  EH, we need to identify the conformal map $z\rightarrow w$ that maps this punctured strip geometry to the logarithmic image of the annulus. As a first step, we perform the map $z\rightarrow \xi=\eE^{\ir\pi z/L}$, that sends the strip to the upper half-plane, as shown in the second panel of Fig.~\ref{fig:map}. After the mapping, the entangling point is sent to $\xi_0=\xi(x_0)$, and the regularized interval $\A_{\epsilon}$ is mapped into the arc $\xi(\A_{\epsilon})$ of unit radius and argument $\theta\in[\frac{\pi}{L}(x_0+\epsilon),L]$.
Then, we map the upper half-plane into the annulus  $\mathbf{D}=\{\zeta, R < |\zeta |\leqslant 1\}$. The conformal transformation $\xi \rightarrow \zeta(\xi) $ is chosen such that the real axis is mapped to the unit circumference of the disk, and the entangling point $\xi_0$ is mapped to the center. In particular, we choose the mapping so that the boundary $\xi(L)$ of the image interval $\xi(\A_{\epsilon})$ is sent to the point $\zeta(\xi(L))=1$ on the unit circumference. This can be achieved via the following function,
\begin{equation}
\zeta(\xi)= \frac{(1+\bar{\xi}_0)}{(1+\xi_0)} \frac{(\xi-\xi_0)}{(\xi-\bar{\xi}_0)}=\frac{\sin(\pi(z-x_0)/2L)}{\sin(\pi(z+x_0)/2L)}.
\end{equation}
The inner radius $R$ of the annulus around the entangling point is then related to the radius $\epsilon$ of the original regularization disk as $R=\zeta(\xi(\epsilon))$,
and the regularized interval is mapped to the segment $\A_R=[R,1]$, as shown in the third panel of Fig.~\ref{fig:map}.
Finally, we can obtain the map to the logarithmic image of the annulus via the  transformation
\begin{equation}
\label{eq:wmap}
w=f(z)=\ln \left[\frac{\sin\frac{\pi(z-x_0)}{2L}}{\sin\frac{\pi(z+x_0)}{2L}} \right],
\end{equation}
such that  $w\in[\ln R,0]\times [0,2\pi)$, and $\mathrm{Im}w=0$ identified with $\mathrm{Im}w=2\pi$. The regularised
interval $\A_{\epsilon}$ is mapped into the interval $[\ln R, 0]$ at $\mathrm{Im} w=0$, that is, on the negative real axis, as shown in the fourth panel of Fig.~\ref{fig:map}.

The conformal transformation \eqref{eq:wmap} can be used to obtain the CFT expression of the EH in the original strip geometry. In particular, after the transformation, the path integral describes a thermal state of a system of length $\ln R$ at inverse temperature $2\pi$. The corresponding EH is
\begin{equation}
    \mathcal{H}= 2 \pi \int_{f(\A_{\epsilon})} T_{00}(w) \dd w.
\end{equation}
The EH in the original geometry can thus be obtained by using the transformation property of the energy-momentum tensor $T(w)=(f'(z))^{-2}T(z)+c/12\{z,w\}$. Taking into account a Jacobian factor $f'(z)$ in the transformation and absorbing the constant term given by the Schwarzian derivative in the normalization of the EH, one finally gets
\begin{equation}
\mathcal{H}=2 \pi \int_\A \beta(x) T_{00}(x)\dd x,
\end{equation}
where 
\begin{equation}
\label{eq:betax}
\beta(x)=\frac{1}{f'(x)}=\frac{L}{\pi}\frac{\cos\left(  \frac{\pi x_0}{L}\right)-\cos\left( \frac{\pi x}{L}\right)}{\sin\left( \frac{\pi x_0}{L}\right)}, \hspace{1cm} x\in[0,L].
\end{equation}
Note that, introducing the shifted variable $y = x-L/2$, the expression \eqref{eq:betax} reproduces the one found in \cite{CT-16} for a symmetric domain $y\in[-L/2,L/2]$.

%%%%%%%%%%%%%%%%%%%%%%%%%%%%%%%%%%%
%%%%%%%%%%%%%%%%%%%%%%%%%%%%%%%%%%%

We now proceed to the inhomogeneous case, considering in particular the free massless Dirac fermion theory in \eqref{eq:inhomDirac}, with the spatial inhomogeneity encoded in the Fermi velocity $v_F(x)$. The theory can be described via the action \cite{DSVC-17}
\begin{equation}
    \mathcal{S}=\frac{1}{2\pi}\int \dd z \dd \Bar{z} \, \eE^{\sigma(x)}\left[ \psi_R^{\dagger} \mytensor{\partial}_{\Bar{z}} \psi_R+ \psi_L^{\dagger} \mytensor{\partial}_z \psi_L\right],
\end{equation}
defined on the Euclidean strip in terms of the isothermal coordinates $(z,\bar z)$. The background metric
\begin{equation}
    \dd s^2=\dd x^2 + v^2_F(x) \dd t^2= \eE^{2 \sigma(x)}\dd z \dd \bar{z}
\end{equation}
is Weyl-equivalent to the flat metric, with the Weyl factor identified as $\eE^{\sigma(x)}=v_F(x)$, and thus %depends only on the spatial variable
the isothermal coordinates must be defined as
\begin{equation}
\label{eq:isothermalx}
    z=\tilde{x}(x)+\ir t, \hspace{1.5 cm} \tilde{x}(x)=\int_{x_{-}}^x \frac{\dd u}{v_F(u)}.
\end{equation}
It is important to emphasize that the integral in
\eqref{eq:isothermalx} is restricted to the domain $u\in[x_{-},x_{+}]$, where the function $v_F(u)$ remains strictly positive. This ensures the transformation is well-defined and the integral converges.
In the transformed coordinates, the width of the Euclidean strip becomes
\begin{equation}
\label{eq:Ltilde}
    \tilde{L}=\tilde{x}(x_+)=\int_{x_{-}}^{x_{+}}\frac{\dd u}{v_F(u)}.
\end{equation}
Additionally, the position of the entangling point is $\tilde{x}_0=\tilde{x}(x_0)$.

A fundamental result in \cite{TRLS-18} is that the expression of the EH in the inhomogeneous background is formally the same as in the homogeneous case, expressed in terms of the new coordinates
\begin{equation}
    \mathcal{H}=2 \pi \int_{\tilde{\A}}\frac{T_{00}(\tilde{x})}{\tilde{f}'(\tilde{x})}\dd \tilde{x},
\end{equation}
where $\tilde{\A}$ is the transformed interval, and $\tilde{f}$ is the function \eqref{eq:wmap} expressed in terms of the tilded quantities. The expression of the EH in the original spatial coordinates can be derived using the relation $T_{00}(\tilde{x})=\tilde{x}'(x)^{-2}T_{00}(x)$, so that
\begin{equation}
    \mathcal{H}= 2\pi \int_{x_0}^{x_+} \frac{T_{00}(x)}{\tilde f'(\tilde x(x))\tilde x'(x)} \, \dd x \, = 2\pi \int_{x_0}^{x_+} \tilde \beta(x) v_F(x) T_{00}(x) \, \dd x \,,
    \label{eq:EHinhom2}
\end{equation}
since $\tilde{x}'(x)=v_F^{-1}(x)$ from \eqref{eq:isothermalx}. In turn, the weight function
\begin{equation}
    \tilde{\beta}(x)=\frac{\tilde{L}}{\pi}\frac{\cos\left(  \frac{\pi \tilde{x}_0}{\tilde{L}}\right)-\cos\left( \frac{\pi \tilde{x}(x)}{\tilde{L}}\right)}{\sin\left( \frac{\pi \tilde{x}_0}{\tilde{L}}\right)}
    \label{eq:betatilde}
\end{equation}
describes a deformation of \eqref{eq:inhomDirac}, i.e.,~it multiplies the inhomogeneous energy density $v_F(x) T_{00}(x)$ and can thus be interpreted as a space-dependent inverse temperature \cite{BE-24}.

Although the CFT calculation yields a strictly local result, this will not be true for the lattice EH \eqref{eq:EHop}. Instead, the operator $\mathcal{T}$ in \eqref{def:THamil} that commutes with the lattice EH, $[\mathcal{T}, \mathcal{H}] =0$, 
includes only nearest-neighbour hoppings. It remains to prove that $\mathcal{T}$ indeed inherits the CFT structure obtained above, and this is done in the next section.

\section{Examples}
\label{sec:examples}

We now apply the CFT treatment of the previous section to free-fermion chains associated to various families of discrete orthogonal polynomials of the Askey scheme, as introduced in Sec. \ref{sec:model}. We show that the elements \eqref{eq:Telements} of the associated commuting tridiagonal matrix can be understood via the discretized version of the inverse temperature \eqref{eq:betatilde}. Besides clarifying the physical origin of the $T$ matrix structure, the CFT approach also fixes the appropriate energy scale, which will be verified in numerical calculations.

\subsection{Krawtchouk chain}
\label{subs:Krawtchouk}

Our first example is the Krawtchouk chain \cite{crampe2019free}. The parameters of the recurrence relation are 
\begin{equation}
    A_n = -p(N-1-n), \qquad C_n= - n(1-p),
\end{equation}
corresponding to the Hamiltonian \eqref{def:Hamil} with
%in the
%Hamiltonian are
%
\begin{equation}
\label{eq:JBKratchouk}
\begin{split}
J_n&=\sqrt{(n+1)(N-1-n)p(1-p)}, \\
B_n&=-(N-1)p-n(1-2p) ,
\end{split}
\end{equation}
and $0<p<1$. The hopping amplitudes are reflection symmetric $J_{N-n-2}=J_n$ and have a semicircular profile, while the potentials are linear with a slope fixed by $p$. The single-particle wavefunctions are then related to the Krawtchouk polynomials via \eqref{phiR}. The corresponding spectrum and dual spectrum are both linear and given by
\begin{equation}
\epsilon_k=k, \qquad
\chi_i=i,
\end{equation}
where $k,i=0,\dots,N-1$. The ground state with $K$ particles is obtained by fixing the chemical potential as
$\mu_0=(\epsilon_{K}+\epsilon_{K+1})/2= K+1/2$.
Substituting into \eqref{eq:Telements}, one arrives at \cite{crampe2019free}
\begin{equation}
\label{eq:TmatKratchouk}
t_i= (i-\ell)J_i, \quad \quad d_i=-(i-1/2-\ell)(B_i+\mu_0).
\end{equation}
The commuting tridiagonal matrix $T$ in \eqref{eq:Ttrid} thus simply corresponds to a linear deformation of the physical Hamiltonian, with the weights of the potentials and hopping amplitudes shifted by a half lattice site.

Our goal is now to reobtain this result for the EH in the CFT framework. Introducing $x=(n+1)/N$, the parameters \eqref{eq:Jmux} in the continuum limit $N\to\infty$ and $Na=1$ are given by
\begin{equation}
%J_n=\frac{1}{2}\sqrt{(n+1)(N-(n+1))}, \quad \quad B_n=K+1-\frac{N}{2},
J(x)=\sqrt{x(1-x)p(1-p)}, \qquad
\mu(x)=\rho-p-x(1-2p),
\end{equation}
where the constant term in the potential is related to the average density $\rho=(K+1)/N$. The Fermi velocity is obtained from \eqref{eq:vF} as
\begin{equation}
v_F(x)=\sqrt{4p(1-p)x(1-x)-\left(\rho-p-x(1-2p)\right)^2},
\end{equation} 
and has a semicircular form. In fact, it is easy to verify that the center and the radius of the semicircle are given by
\begin{equation}
\bar{x}=\frac{1}{2}[1-(1-2\rho)(1-2p)], \qquad
R=2\sqrt{p(1-p)\rho(1-\rho)},
\end{equation} 
respectively, such that the Fermi velocity can be rewritten as
\begin{equation}
v_F(x)=\sqrt{R^2-\left(x-\bar x \right)^2}.
\label{eq:vFK}
\end{equation} 
In particular, the center is in the middle of the chain ($\bar x =1/2$) for half filling and arbitrary $p$, or $p=1/2$ and arbitrary filling. Note, however, that the fermionic density itself has reflection symmetry only at $p=1/2$. The Fermi velocity is also symmetric under the simultaneous exchanges $\rho \to 1-\rho$ and $p \to 1-p$, which corresponds to a particle-hole transformation.

The curved-space CFT corresponding to \eqref{eq:vFK} is defined on the segment $[x_-,x_+]$, where $x_\pm = \bar x \pm R$ are the zeros of the Fermi velocity, $v_F(x_\pm)=0$. The isothermal coordinates \eqref{eq:isothermalx} are obtained as
\begin{equation}
    \tilde{x}(x)=\int_{x_-}^x\frac{\dd u}{v_F(u)}= \pi- \arccos \left(\frac{x-\bar x}{R}\right),
\end{equation}
and the length of the strip is given by $\tilde L =\tilde x(x_+)=\pi$. Furthermore, it is easy to see that
\begin{equation}
\label{eq:sinx0}
\sin(\frac{\pi \tilde x_0}{\tilde L})=\frac{v_F(x_0)}{R}.
\end{equation}
Plugging these expressions into \eqref{eq:betatilde}, we finally obtain the CFT expression of the inverse temperature,
\begin{equation}
\label{eq:tildebetaK}
\tilde \beta(x) = \frac{x-x_0}{v_F(x_0)}.
%      \mathcal{H}=\frac{2\pi}{v_F(x_0)} \int_{A}(y-y_0) H(y).
\end{equation}

\begin{figure}[t]
         \begin{subfigure}[b]{0.46\textwidth}
         \centering
    \includegraphics[width=\textwidth]{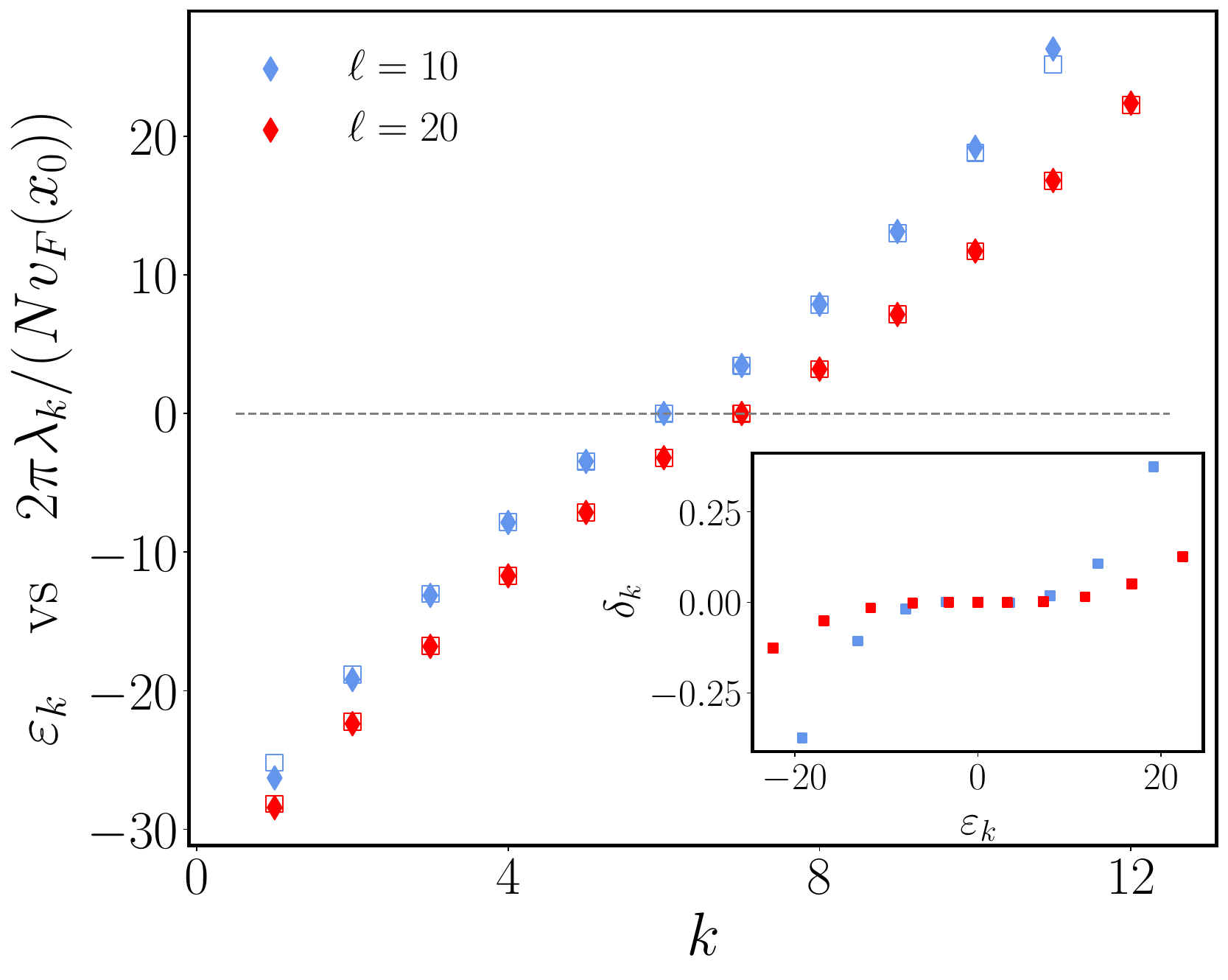} 
     \end{subfigure}
     \hfill
     \begin{subfigure}[b]{0.46\textwidth}
    \includegraphics[width=\textwidth]{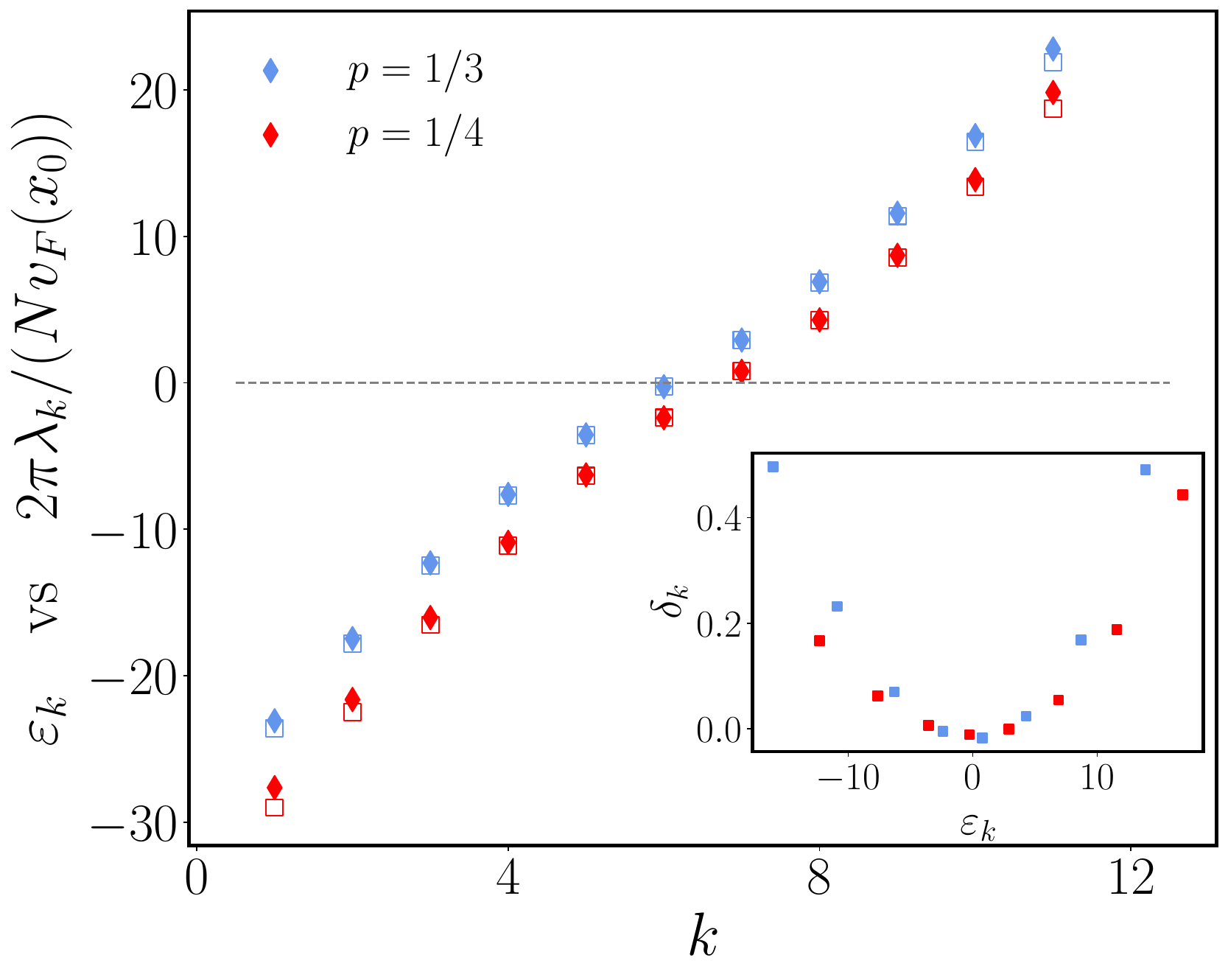} 
     \end{subfigure}
     \caption{Comparison between the entanglement spectrum $\varepsilon_k$ (full symbols) and the scaled eigenvalues $2\pi\lambda_k/(N v_F(x_0))$ of the $T_{\mathcal{A}}$ matrix (empty symbols) for the Krawtchouck chain in the symmetric $p=1/2$ case at different  $\ell$ (left), and for $p=1/4,1/3$ with fixed $\ell=20$ (right). The data are obtained numerically for $N=60$ and at half filling. The insets show the corresponding difference $\delta_k=\varepsilon_k-2\pi \lambda_k/(N v_F(x_0))$ between the two spectra, as function of $\varepsilon_k$.}
     \label{fig:specKrawt}
\end{figure}

Hence, the EH precisely has the BW form, and is in complete agreement with the result \eqref{eq:TmatKratchouk} for the $T$ matrix. Moreover, while the commutation relation $[T_\A,h]=0$ only implies a common set of eigenvectors, the CFT calculation suggests that the two matrices are proportional with a scale factor fixed via the Fermi velocity \eqref{eq:vFK} at the boundary of the subsystem. Indeed, combining Eqs.~\eqref{eq:inhomDirac}, \eqref{eq:EHinhom2}, \eqref{eq:TmatKratchouk} and \eqref{eq:tildebetaK}, and setting $x_0=(\ell+1)/N$ and $x-x_0=(i-\ell)/N$, one has the prediction
\begin{equation}
h \simeq \frac{2 \pi}{N v_F(x_0)} T_\A,
\end{equation}
which can be checked in numerical calculations, as shown in Fig.~\ref{fig:specKrawt}. In general, the agreement is very good between the low-lying parts of the respective spectra, however, some deviations are visible for larger eigenvalues. This is completely analogous to what has been observed in previous studies of lattice models with a commuting $T_\A$ matrix  \cite{EP-17,BE-24}, where the EH was written as a power series
\begin{equation}
\label{eq:EHBW}
    h=\sum_{m=1}^{\infty} \alpha_m T_\A^m.
\end{equation}
In particular, the BW form corresponds to setting $\alpha_1=2\pi/N v_F(x_0)$, while the deviations are characterized by the coefficients $\alpha_m$ with $m>1$. These deviations are shown in the insets of Fig.~\ref{fig:specKrawt}, suggesting that the lowest nonvanishing order is given by $m=3$ in the particle-hole symmetric case $p=1/2$ and $\rho=1/2$ (left) while for general $p$ (right) one has also quadratic ($m=2$) corrections. The higher powers of $T_\A$ will generate longer range hopping terms in the lattice EH \eqref{eq:EH}, which can only be removed by a proper continuum limit \cite{RSC-22}. Nevertheless, as one can deduce %obvious 
from Fig.~\ref{fig:specKrawt}, the BW approximation should already give a good 
description of 
physical quantities which are not sensitive to the higher part of the spectrum, such as 
the entanglement entropy.

%%%%%%%%%%%%%%%%%%%%%%%%%%%%%%%%%

%

%%%%%%%%%%%%%%%%%%%%%%%%%%%%%%%%%%%%%%%

\subsection{Dual Hahn chain}
\label{subs:dHahn}

Our second example is the dual Hahn chain with parameters
\begin{equation}
    A_n=-(N-1-n)(n+\gamma+1),\qquad C_n=-n(N-n+\delta),
\end{equation}
corresponding to
\begin{equation}
\begin{split}
J_n&=\sqrt{(n+1)(N-1-n)(n+\gamma+1)(N-1-n+\delta)},\\
B_n&=-(N-1-n)(n+\gamma+1)-n(N-n+\delta)
\end{split}
\label{eq:JBdualHahn}
\end{equation}
where $\gamma,\delta>-1$.
From the recurrence and difference relations,
one obtains \cite{koekoek2010hypergeometric}
\begin{equation}
\label{eq:specdualHahn}
\epsilon_k=k(k+1+\gamma+\delta), \qquad
\chi_i=i. %\qquad k,i=0,\cdots,N-1.
\end{equation}
Hence the spectrum is now quadratic, while the dual spectrum remains linear as for the Krawtchouk chain. Since the latter one uniquely determines the deformation of the Hamiltonian, the result \eqref{eq:TmatKratchouk} for the $T_{\mathcal{A}}$ matrix elements follows immediately, with the corresponding parameters given by \eqref{eq:JBdualHahn}.

We now move on to the continuum limit.
Introducing the continuous scaling variables $x=(n+1)/N$, $x_1=\gamma/N$ and $x_2=\delta/N$, in the limit $N\to \infty$ one obtains
\begin{equation}
\begin{split}
J(x)&=-\sqrt{x(1-x)(x+x_1)(1-x+x_2)},\\
\mu(x) &= -2x(1-x)-x_1(1-x)-x_2 x + \rho(\rho+x_1+x_2).
\end{split}
\end{equation}
Note that the lattice parameters \eqref{eq:JBdualHahn} are now scaled by a factor of $N^2$ in order to have a well defined continuum limit. Compared to the Krawtchouk case \eqref{eq:JBKratchouk}, the hopping is more complicated and does not have a semicircular form. However, it is direct to see that the quartic terms in $4J^2(x)$ are exactly canceled by the corresponding ones in $\mu^2(x)$ in the expression \eqref{eq:vF}. The Fermi velocity thus preserves its semicircular form, and with simple manipulations one can bring it to the form
\begin{equation}
\label{eq:vFdH}
\begin{split}
v_F(x)= \xi \sqrt{R^2 - \left(x-\bar x\right)^2},
\end{split}
\end{equation}
with the parameters given by
\begin{equation}
\begin{split}    
    &\xi=x_1+x_2+2\rho,\\
    &\bar x = \frac{1}{2}+\frac{(x_2-x_1)((\rho-1/2)(x_1+x_2)+\rho^2)}{\xi^2}, \\
    &R^2 = \frac{4\rho(1-\rho)(\rho+x_1)(\rho+x_2)(\rho+x_1+x_2)(1+\rho+x_1+x_2)}{\xi^4}.
\end{split}
\end{equation}

The structure of $v_F(x)$ thus differs from the Krawtchouk case \eqref{eq:vFK} only by an overall factor $\xi$, which changes the length scale in the coordinate transformation \eqref{eq:isothermalx}. Nevertheless, this scale enters trivially, as the arguments of the trigonometric functions depend only on the ratios of lengths in the inverse temperature \eqref{eq:betatilde}.
%, 
In turn, one arrives at the exact same result as in \eqref{eq:tildebetaK}, which depends on the dual Hahn parameters only via the corresponding Fermi velocity in \eqref{eq:vFdH}. The setup is thus completely analogous to the Krawtchouk case, and yields a BW form for the EH. A comparison of the entanglement spectrum to the appropriately rescaled $T_\A$ matrix spectrum (not shown) delivers very similar results as in Fig.~\ref{fig:specKrawt}.

\subsection{Hahn chain}
\label{subs:Hahn}

Next we consider the chain associated to Hahn polynomials, which will turn out to be more complicated. The coefficients of the recurrence relation read
\begin{equation}
\label{eq:ACHahn}
A_n=-\frac{(n+\alpha+\beta+1)(n+\alpha+1)(N-1-n)}{(2n+\alpha+\beta+1)(2n+\alpha+\beta+2)}, \qquad C_n=-\frac{n(n+\alpha+\beta+N)(n+\beta)}{(2n+\alpha+\beta)(2n+\alpha+\beta+1)},
\end{equation}
with the parameters restricted to $\alpha,\beta\footnote{Here, $\beta$ differs from the inverse temperature $\beta(x)$ introduced in the previous sections. We have retained the standard notation for consistency with both the reference \cite{koekoek2010hypergeometric} and the broader literature.}>-1$,
and the corresponding spectral relations are given by
\begin{equation}
\epsilon_k = k, \qquad
\chi_i = i(i+\alpha+\beta+1).% \qquad k,i=0,\cdots,N-1.
\end{equation}
The coefficients $J_n$ and $B_n$ can be obtained by plugging $A_n,C_n$ from \eqref{eq:ACHahn} in \eqref{def:JnBn}, but their exact expressions are cumbersome and we do not reproduce them here.
Compared to the dual Hahn case, the role of the spectrum and dual spectrum are simply interchanged, which explains the nomenclature. Furthermore, since the dual spectrum is now quadratic, we find using~\eqref{eq:Telements} that the commuting $T_{\mathcal{A}}$ matrix now corresponds to a parabolic deformation with elements
\begin{equation}
\label{eq:THahn}
    \begin{split}
        t_i&=(i-\ell)(i+\ell+\alpha+\beta+2)J_i,\\
        d_i&= -[(i-1/2-\ell)(i-1/2+\ell+\alpha+\beta+2)-1/4](B_i+\mu_0).
    \end{split}
\end{equation}
The first root of the parabola is located at the boundary of the subsystem, while the second root is negative and thus lies outside of the chain.

\begin{figure}[t]
\centering 
\includegraphics[width=0.6 \textwidth]{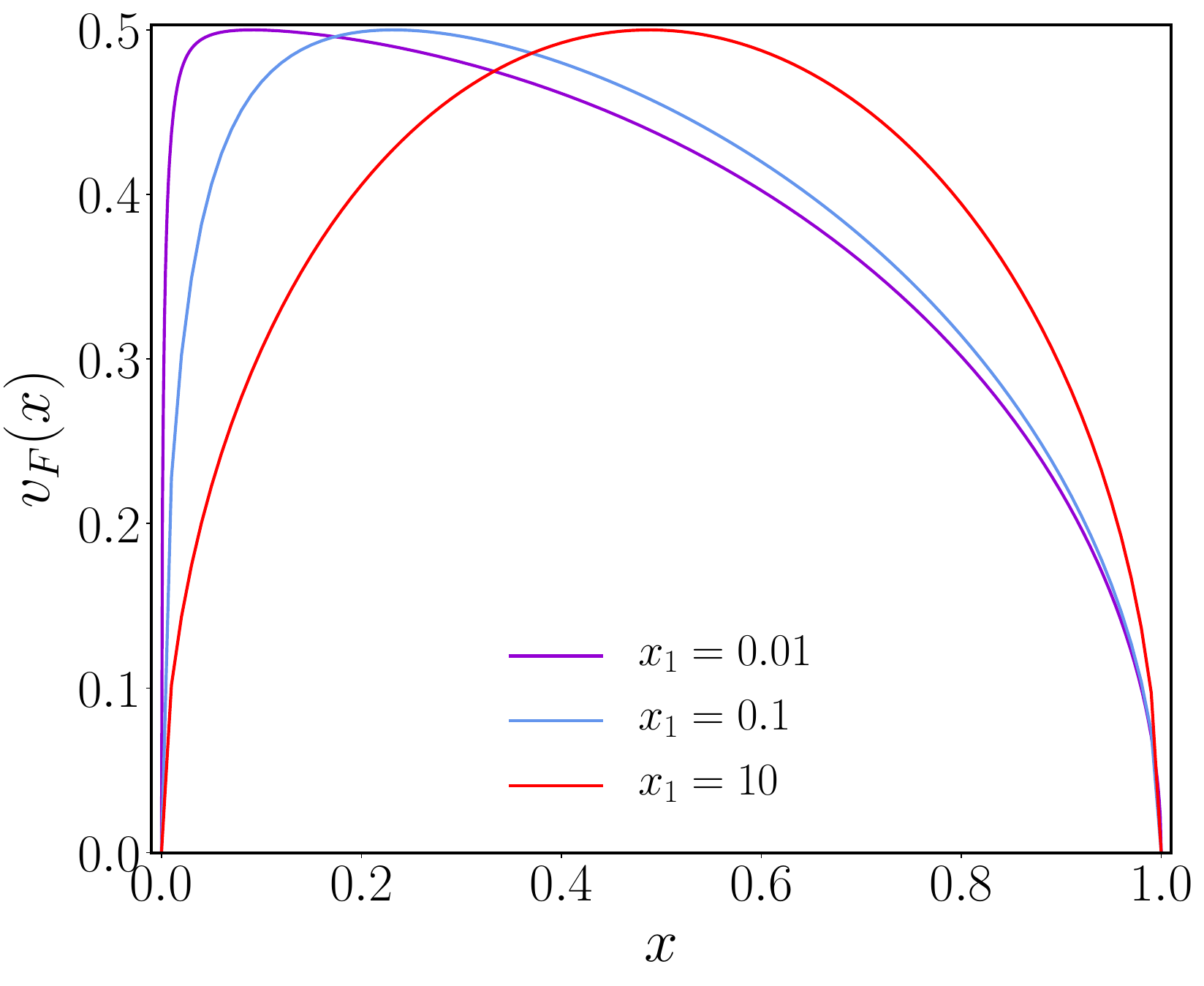} 
     \caption{Fermi velocity \eqref{eq:vFHahn} of the half-filled Hahn chain for different values of $x_1$.}
     \label{fig:vFHahn}
\end{figure}

We now show that the same parabolic deformation appears also for the EH in the continuum limit. For the clarity of the presentation, we first focus on the symmetric case $\alpha=\beta$, which yields a homogeneous chemical potential $\mu(x)=\rho-1/2$. A further simplification occurs when we restrict ourselves to half filling $\rho=1/2$, where from \eqref{eq:qF} one finds a homogeneous density. Setting $x_1=\alpha/N$, one obtains
\begin{equation}
\label{eq:vFHahn}
v_F(x)=2J(x)=\sqrt{\frac{x(1-x)(x+2x_1)(1+x+2x_1)}{(2x+2x_1)^2}}.
\end{equation}
It is clear from the above expression that, in general, the Fermi velocity does not possess a semicircular form, which can only be recovered in the limit $x_1 \to \infty$. Indeed, the Hahn polynomials in the scaling limit $\alpha=pt$ and $ \beta=(1-p)t$ with $t\rightarrow \infty$ reproduce the Krawtchouk polynomials \cite{koekoek2010hypergeometric}. This explains the
observation that \eqref{eq:vFHahn} converges to the expression of $v_F(x)$ for the Krawtchouck chain \eqref{eq:vFK} in the symmetric case $p=1/2$.
In the opposite limit $x_1 \to 0$, instead, one obtains a quarter circle $v_F(x) \to \sqrt{1-x^2}/2$. For intermediate values of $x_1$, the shape of the Fermi velocity interpolates between these two limiting forms, as shown in Fig.~\ref{fig:vFHahn}, becoming more asymmetric for decreasing values of $x_1 \rightarrow 0$.
%

%%%%%%%%%%%%%%%%%%%%%%%%%%%%%%%%%%%
%%%%%%%%%%%%%%%%%%%%%%%%%%%%%%%%%%%

We now proceed to the evaluation of the inverse temperature, for which one first needs to integrate the inverse of $v_F(x)$ in \eqref{eq:vFHahn}. It can be done by brute force, splitting the integral in two pieces and finding closed form expressions in terms of elliptic integrals. The details of this calculation 
can be found in Appendix \ref{sec:appA}. However, adding the contributions and applying some identities, the result can eventually be brought into a very simple form and reads
\begin{equation}
\label{eq:tildexHahn}
    \tilde{x}(x)= \arccos\left[1-\frac{2x(x+2x_1)}{1+2x_1}\right].
\end{equation}
Plugging this result into \eqref{eq:betatilde}, the inverse temperature has the expected parabolic form
\begin{equation}
\label{eq:tildebetaHahnsym}
    \tilde{\beta}(x)=\frac{(x-x_0)(x+x_0+2x_1)}{(2x_0+2x_1)v_F(x_0)},
\end{equation}
whose discretization agrees perfectly with the results for the $T_{\mathcal{A}}$ matrix elements in \eqref{eq:THahn}. Similarly to the linear case \eqref{eq:tildebetaK}, the CFT calculation fixes the coefficient governing the amplitude of the deformation,  which is now given by the Fermi velocity multiplied with the distance between the zeroes of the parabola. This suggests that the low-lying entanglement spectrum can be approximated via the $T_\A$ matrix eigenvalues $\lambda_k$ as
\begin{equation}
\label{eq:epslamHahn}
\varepsilon_k \simeq \frac{2\pi}{(2x_0+2x_1)v_F(x_0)}
\frac{\lambda_k}{N^2},
\end{equation}
where $v_F(x)$ is given by \eqref{eq:vFHahn}, and the $N^2$ factor is needed to bring the parabolic deformation in the matrix elements \eqref{eq:THahn} into a scaling form.

%%%%%%%%%%%%%%%%%%%%%%%%%%%%%%%%%%%
\begin{figure}[t]
\centering 
\includegraphics[width=0.6 \textwidth]{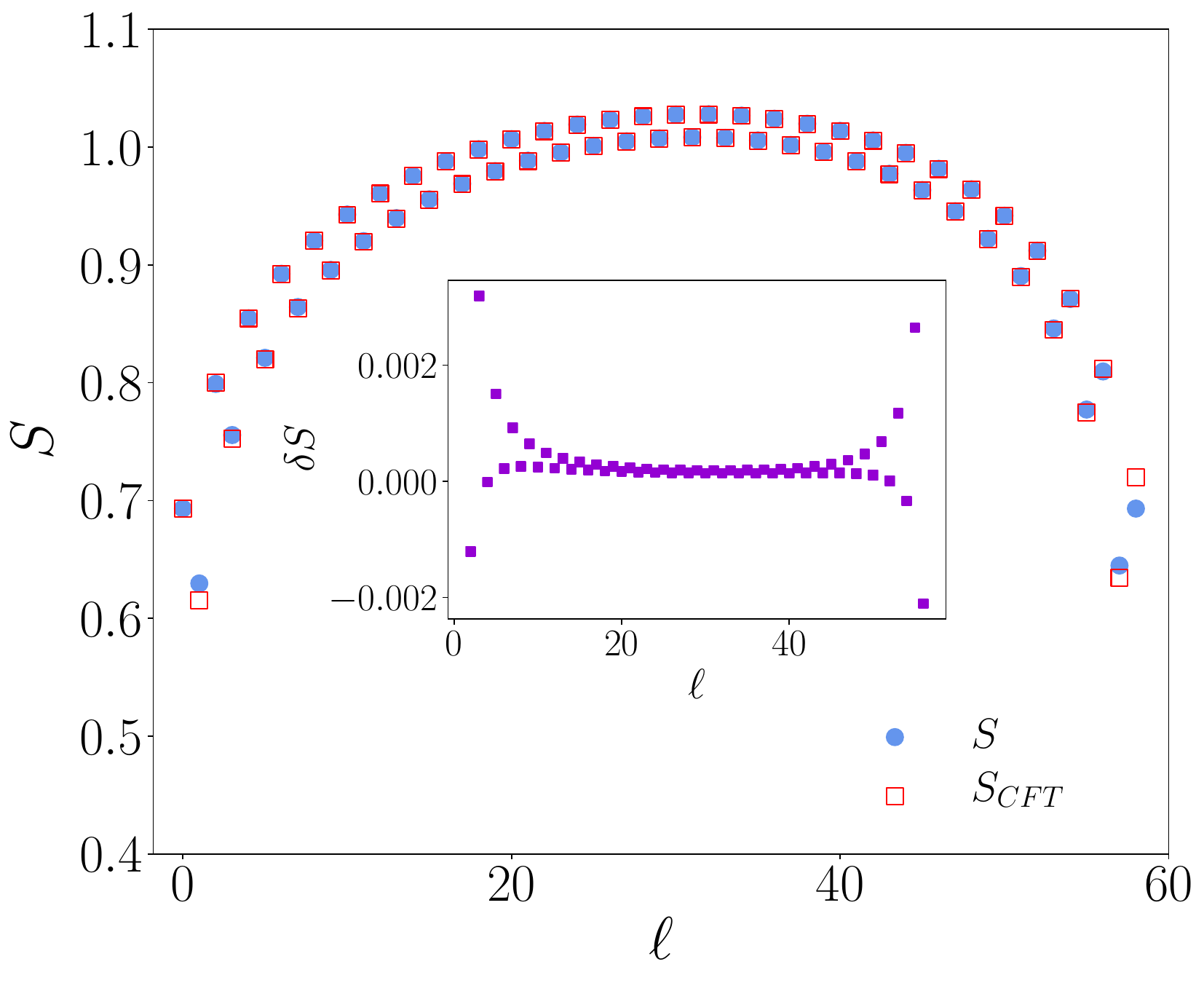} 
     \caption{Comparison of the entropies $S$ and $S_{CFT}$ for the Hahn chain as function of $\ell$, computed from \eqref{eq:S} using the eigenvalues $\varepsilon_k$ and $2\pi \lambda_k/ (N^2 (2x_0+2x_1)v_F(x_0))$, respectively. The data are obtained at half filling with $N=60$ and $x_1=x_2=8/N$. The inset shows the difference $\delta S=S-S_{CFT}$.}
     \label{fig:SHahn}
\end{figure}
%%%%%%%%%%%%%%%%%%%%%%%%%%%%%%%%%%%

In order to test the relation between the spectra, we calculate the entanglement entropy
\begin{equation}
\label{eq:S}
    S=\sum_k \left(\frac{\varepsilon_k}{\eE^{\varepsilon_k}+1}+\ln(1+\eE^{-\varepsilon_k})\right),
\end{equation}
and compare it to the approximation $S_{CFT}$, obtained %computed 
by replacing the eigenvalues $\varepsilon_k$ by the r.h.s of \eqref{eq:epslamHahn} in \eqref{eq:S}. The results are shown in Fig. \ref{fig:SHahn}, where the values $S$ (full symbols) and $S_{CFT}$ (empty symbols) are compared as a function of the boundary position $\ell$ for a chain of length $N=60$. The agreement between the two quantities is very good, with slight deviations when $\ell$ is close to the boundaries of the chain. This behavior is highlighted in the inset of Fig.~\ref{fig:SHahn}, where we show the difference $\delta S= S-S_{CFT}$ as function of $\ell$.

Finally, we note that the generic case with $x_1\neq x_2=\beta/N$ and arbitrary densities $\rho$ can also be dealt with by reverse engineering. Namely, one can easily guess from the form of the $T_{\mathcal{A}}$ matrix \eqref{eq:THahn}, that the general solution for $\tilde \beta(x)$ should follow by replacing $2x_1 \to x_1 + x_2$ in \eqref{eq:tildebetaHahnsym}. The question is then, whether we can construct the function $\tilde x(x)$, which delivers this result. To this end, let us observe that the denominator of \eqref{eq:tildebetaHahnsym} should originate from the sine factor in the CFT formula \eqref{eq:betatilde}, up to a possible multiplicative constant. This is indeed the case in our calculation for $x_1=x_2$ and $\rho=1/2$. Since the relation must hold for \emph{arbitrary} values of $x_0$, we obtain the condition
\begin{equation}
\label{eq:sinx}
\sin(\frac{\pi \tilde x(x)}{\tilde L})=\frac{(2x+x_1+x_2)v_F(x)}{R^2},
\end{equation}
which is simply the analog of the relation \eqref{eq:sinx0} for the case of parabolic deformations.
Note that, since one has $\tilde x(x_+)=\tilde L$ and $v_F(x_+)=0$, the boundary condition is automatically satisfied. Furthermore, the scale factor can be fixed as the maximum of the function $(2x+x_1+x_2)v_F(x)=\sqrt{f(x)}$, which is given by the square root of a fourth-order polynomial $f(x)$, see Appendix \ref{sec:appA} for details. A simple calculation shows that there is a single maximum in $x\in[x_-,x_+]$ and its value can be found as
\begin{equation}
R^2 = 2\sqrt{\rho(1-\rho)(\rho+x_1)(1-\rho+x_2)}.
\end{equation}
Having fixed our ansatz, the last step is to take the derivative of \eqref{eq:sinx} and impose the relation $\tilde x'(x)=1/v_F(x)$,
which follows from the definition \eqref{eq:isothermalx}. This yields a nontrivial condition for the function $f(x)$, which can be shown to be satisfied, concluding our proof for $\tilde \beta(x)$.

\subsection{Racah chain}
\label{subs:Racah}

\begin{figure}
    \centering
    \includegraphics[width=0.6\linewidth]{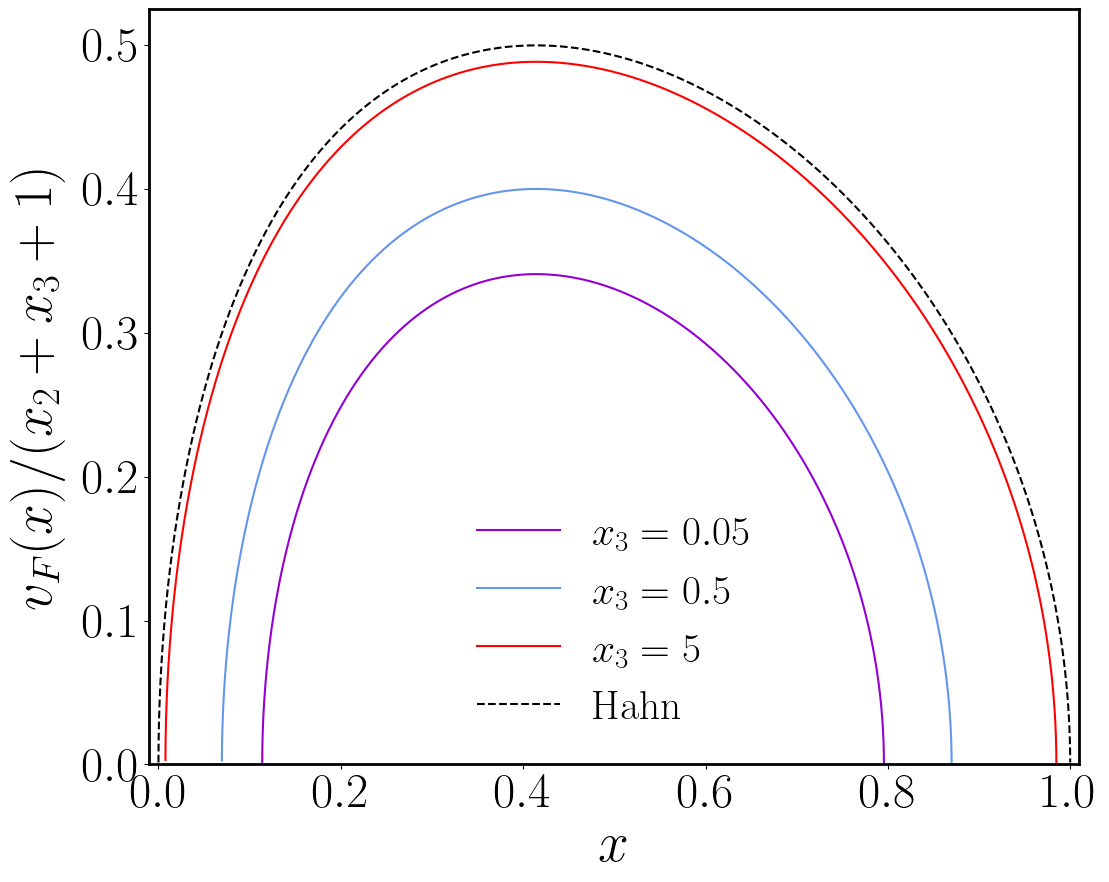}
    \caption{Fermi velocity of the half-filled Racah chain with $x_1 = 3$ and $x_2 = 1$, for different values of $x_3$. This chain reduces to the Hahn case with $x_1 = x_2 =1$ in the $x_3 \rightarrow \infty$ limit.  }
\label{fig:vfracah}
\end{figure}

The final case we consider is the chain associated to Racah polynomials. They constitute the most general family of classical discrete orthogonal polynomials in the Askey scheme, and have recurrence coefficients given in terms of parameter $\alpha$, $\beta$, $\gamma$ and $\delta$ as 
\begin{equation}\label{def:AnRacah}
\begin{split}
    A_n &= \frac{(\alpha +n+1) (\gamma +n+1) (\alpha +\beta +n+1) (\beta +\delta +n+1)}{(\alpha +\beta +2 n+1) (\alpha +\beta +2 n+2)},\\
C_n &= \frac{n (\beta +n) (\alpha -\delta +n) (\alpha +\beta -\gamma +n)}{(\alpha +\beta +2 n+1) (\alpha +\beta +2 n)}.
\end{split}
\end{equation}
Similarly as for the Hahn chain, the coefficients $J_n$ and $B_n$ can be obtained from \eqref{def:JnBn}, but their exact expressions are cumbersome and we do not reproduce them here.
The parameters associated to these polynomials must satisfy a truncation condition, which we take as $\alpha = -N$ in the following. For the Racah polynomials both spectra are quadratic,
\begin{equation}
    \epsilon_k = k(k+\gamma+\delta+1), \qquad
    \chi_i = i(i+\alpha+\beta+1),
\end{equation}
and the functional form of the Fermi velocity is similar to the Hahn case with
\begin{equation}
\label{eq:vFR}
    v_F(x) = \sqrt{\frac{g(x)}{(2x+x_1-1)^2}},
\end{equation}
where $g(x)$ is a fourth-order polynomial, depending on the scaling variables $x_1=\beta/N$, $x_2=\gamma/N$, $x_3=\delta/N$ and the density $\rho$. The full expression of $g(x)$ is rather lengthy and we report it only in Appendix \ref{app:racah}. Note that, despite the singularity in \eqref{eq:vFR}, one can easily verify that both positive roots of $g(x)$ satisfy $x_\pm>(1-x_1)/2$, and thus $v_F(x)$ is well defined and regular for all values of $x_1$. Its behaviour for some fixed $x_1,x_2$ and increasing values of $x_3$ at half filling is shown in Fig.~\ref{fig:vfracah}. Note that the curves are rescaled to allow for a better comparison. In particular, $x_3 \to \infty$ corresponds to a special Racah $\to$ Hahn limit \cite{koekoek2010hypergeometric}, as illustrated by the dashed line.

Since the direct evaluation of the integral \eqref{eq:isothermalx} would be too cumbersome, we follow again a different route to identify the inverse temperature. Namely, we shall recast the expression \eqref{eq:vFR} of the Fermi velocity in terms of $\tilde{\beta}(x)$ by inverting the relation \eqref{eq:betatilde}. Indeed, the space-dependent inverse temperature and its derivative are expressed as
\begin{equation}\label{eq:cond1v2}
    \tilde{\beta}(x)= \frac{1}{\Omega}\left(\Phi - \sqrt{1 + \Phi^2} \cos\left(\Omega \tilde{x}(x)\right)\right), \qquad  \tilde{\beta}'(x)  = \frac{\sqrt{1 + \Phi^2} \sin\left( \Omega \tilde{x}(x) \right)}{ v_F(x)},
\end{equation}
where we defined $\Omega = \pi/\tilde{L}$ and $\Phi = \cot\left(  \frac{\pi \tilde{x}_0}{\tilde{L}}\right)$. Isolating $v_F(x)$ from these equations yields
\begin{equation}\label{eq:prop2}
          v_F(x)  = \frac{ \sqrt{1 - \Omega^2 \tilde{\beta}^2(x) + 2 \Phi  \Omega  \tilde{\beta}(x)} }{\tilde{\beta}'(x) }.
    \end{equation}

One should stress that the above relation must hold for arbitrary inhomogeneities. In particular, one can see that a linear $\tilde\beta(x)$ must necessarily originate from a semicircular $v_F(x)$, as we have observed in the Krawtchouk and dual Hahn examples. For a Fermi velocity $v_F(x)$ of more complicated form, the inverse temperature can be determined by identifying the function $\tilde{\beta}(x)$ so that Eq.~\eqref{eq:prop2} holds. In particular, the form \eqref{eq:vFR} with a fourth-order polynomial $g(x)$ immediately suggests to use a parabolic $\tilde\beta(x)$ in
Eq.~\eqref{eq:prop2}. In fact, it is straightforward to check that the local inverse temperature 
associated to the Racah chain can be expressed as
\begin{equation}
    \tilde{\beta}(x) = \frac{(x-x_0)(x+x_0+x_1-1)}{(2 x_0 + x_1 - 1) v_F(x_0)},
\end{equation}
with coefficients $\Omega$ and $\Phi$  provided in Appendix \ref{app:racah}. The result agrees perfectly with the $T_{\mathcal{A}}$ matrix structure that is identical to the Hahn case \eqref{eq:THahn} after the substitution $\alpha=-N$. The entanglement spectrum can thus be matched as
\begin{equation}
\label{eq:epslamRacah}
\varepsilon_k \simeq \frac{2\pi}{(2x_0+x_1-1)v_F(x_0)}
\frac{\lambda_k}{N^2}.
\end{equation}
A numerical comparison of the above spectra was performed for various system sizes and parameters (not shown), revealing a level of agreement similar to that illustrated in Fig.~\ref{fig:specKrawt} for the Krawtchouk.

\section{Discussion}
\label{sec:out}

We have studied hopping chains with particular forms of inhomogeneity, where the single-particle eigenstates are related to discrete orthogonal polynomials, satisfying a three-term difference relation in addition to a three-term recurrence relation. Due to the bispectrality of these functions, one can immediately write down a tridiagonal matrix $T_\A$ that commutes with the ground-state EH of the chain, and corresponds to a spatial deformation of the physical Hamiltonian. In particular, the deformation is entirely determined by the dual spectrum, and hence is either linear or parabolic for the various families of discrete orthogonal polynomials we considered. We showed that this weight factor can be identified with the local inverse temperature that follows from a CFT treatment of the problem in the continuum limit. This procedure automatically yields the appropriate energy scale for the EH, and we have verified that the correspondingly rescaled $T_\A$ matrix eigenvalues give an excellent approximation of the entanglement spectrum.

Our work opens a number of interesting questions. It is a striking observation that the dual spectrum of orthogonal polynomials satisfying a simple difference relation can be matched with the inverse temperature in the EH. However, it remains to be understood why they are restricted to linear or parabolic deformations. It would be intriguing to explore whether a connection between more general $\tilde \beta(x)$ and specific properties of orthogonal polynomials extends beyond the Askey scheme. On the other hand, one could try to find further examples of hopping chains with bispectral properties by engineering the couplings, such that they lead to a semicircular $v_F(x)$. The simplest example is the so-called gradient chain \cite{BE-24}, with constant $J_n$ and linear potential $B_n$, where the solutions for an infinite chain are Bessel functions and a commuting $T$ matrix indeed exists \cite{BOO-00}.

Another natural extension of our studies would be to consider $q$-analogues of polynomials in the Askey scheme, and the associated hopping chains. Fermionic chains associated to $q$-Racah polynomials, representing the most general case, have been shown to admit a commuting 
$T$-matrix that can be diagonalized using a method based on algebraic Bethe ansatz \cite{bernard2023computation}. It would be interesting to see how the calculations described here generalize to this case and what is the inverse temperature it yields. Furthermore, one could also explore orthogonal polynomials of the Askey scheme with a continuous weight, which can be associated to a nonrelativistic Fermi gas in an inhomogeneous potential. In particular, the simplest case of Hermite polynomials corresponds to the harmonic trap, which has recently been shown to yield a BW form of the EH \cite{BE-24}. The role of the $T$ matrix is then played by a second-order differential operator which commutes with the correlation kernel, and was shown to exist for all the families of classical orthogonal polynomials \cite{GRUNBAUM1983491}. A systematic study of these cases and the associated inverse temperatures need to be addressed in future work. Finally, there exist other families of solvable inhomogeneous models related to orthogonal polynomials, and whose continuum limit is described by curved-space CFT \cite{FA20,FA21}. It could be worth investigating the EH of these models to see how our results generalize to these cases.

\section*{Acknowledgement}
This research was funded in part by the Austrian Science Fund (FWF) Grant-DOI: 10.55776/P35434. For the purpose of open access, the authors have applied a CC BY public copyright licence to any Author Accepted Manuscript version arising from this submission. RB and VE are grateful for the hospitality of the Centre de Recherches Math\'ematiques, Universit\'e de Montr\'eal, during the initial phase of this work.
GP held FRQNT and CRM-ISM postdoctoral fellowships, and received support from the Mathematical Physics Laboratory of the CRM while this work was carried out. PAB holds an Alexander-Graham-Bell scholarship from the Natural Sciences and Engineering Research Council of Canada (NSERC).  
The research of LV is funded in part by a Discovery Grant from NSERC.

\appendix

\section{Calculation of $\tilde{\beta}(x)$ for the Hahn chain}
\label{sec:appA}

In this appendix, we show how to compute \eqref{eq:tildebetaHahnsym}. We start by evaluating the isothermal coordinates \eqref{eq:isothermalx}. Substituting \eqref{eq:vFHahn}, the expression can be split in the sum of two integrals. For the first integral, we can use the formula (BY 256.00) in \cite{Gradshteyn}
\begin{equation}
    \int_0^x \frac{\dd x'}{\sqrt{(a-x')(x'-b)(x'-c)(x'-d)}}= \frac{2}{\sqrt{(a-c)(b-d)}}F(\lambda,r),
\end{equation}
with $a\geqslant x >b >c >d $,  where $F(\varphi,k)$ is the incomplete elliptic integral of the first kind,
\begin{equation}
    F(\lambda,r)=\int_0^{\lambda}\frac{\dd x'}{\sqrt{1-r^2\sin^2(x')}}
\end{equation}
and the parameters are defined as
\begin{equation}
    \lambda=\arcsin\sqrt{\frac{(a-c)(x-b)}{(a-b)(x-c)}}, \hspace{1cm} r=\sqrt{\frac{(a-b)(c-d)}{(a-c)(b-d)}}.
\end{equation}
 In particular, we have for the roots $a=1,b=0,c=-2x_1$ and $d=-2x_1-1$, so that
 \begin{equation}
     \label{eq:lambdar}\lambda=\arcsin\sqrt{\frac{(1+2x_1)x}{x+2x_1}}, \hspace{1.5 cm} r=\frac{1}{1+2x_1}.
 \end{equation}

For the second integral, we can use the formula (BY 256.11) in \cite{Gradshteyn},
\begin{equation}
    \int_0^x \frac{x'\dd x'}{\sqrt{(a-x')(x'-b)(x'-c)(x'-d)}}=\frac{2}{\sqrt{(a-c)(b-d)}}\left[ (b-c)\Pi\left( \lambda, \frac{a-b}{a-c},r\right)+cF(\lambda,r) \right],
\end{equation}
where $\Pi$ denotes the incomplete elliptic integral of the third kind, defined as
\begin{equation}
    \Pi(\phi,n^2,k)=\int_0^{\phi}\frac{\dd x'}{(1-n^2\sin^2x')\sqrt{1-k^2\sin^2x}}.
\end{equation}
Adding the two contributions and substituting for the roots we find
\begin{equation}
\label{eq:tildexHahnapp}
        \tilde{x}(x)=
\frac{4}{1+2x_1}\left[
x_1 F(\lambda,r) + 2x_1\Pi\left(\lambda,r,r\right) - 2x_1 F(\lambda,r)\right]=
2(1-r)\left[2\Pi\left(\lambda,r,r\right) - F(\lambda,r)\right].
   %\end{split}
\end{equation}
This expression can be further simplified by using the identity (17.7.22) in \cite{abramowitz_stegun},
\begin{equation}
    (1-r)[2 \Pi(\lambda,r,r)-F(\lambda,r)]=\arctan[(1-r)\tan(\lambda)/\Delta(\lambda)], \hspace{1cm} \Delta(\lambda)=\sqrt{1-r^2\sin^2\lambda}.
\end{equation}
Using \eqref{eq:lambdar} and the identity $\tan(\arcsin z)=z/\sqrt{1-z^2}$, we have
\begin{equation}
    \Delta(\lambda)=\sqrt{\frac{2x_1(1+x+2x_1)}{(1+2x_1)(x+2x_1)}}, \hspace{1cm} \tan(\lambda)=\sqrt{\frac{x(1+2x_1)}{2x_1(1-x)}}.
\end{equation}
Using $\tilde{L}=\tilde{x}(1)$ with $\lambda(1)=\pi/2$, and substituting in \eqref{eq:tildexHahn} we have
\begin{equation}
    \frac{\tilde{x}}{\tilde{L}}=\frac{2 \Pi(\lambda,r,r)-F(\lambda,r)}{2 \Pi(\pi/2,r,r)-F(\pi/2,r)}=\frac{2}{\pi}\arctan\sqrt{\frac{x(x+2x_1)}{(1-x)(1+x+2x_1)}}.
\end{equation}
Finally, using the identity $\cos (2 \arctan z)=(1-z^2)/(1+z^2)$, we find
\begin{equation}
    \cos (\frac{\pi \tilde{x}}{\tilde{L}})= \frac{(1-x)(1+x+2x_1)-x(x+2x_1)}{(1-x)(1+x+2x_1)+x(x+2x_1)}=1-\frac{2x(x+2x_1)}{1+2x_1}.
\end{equation}
Plugging this expression into \eqref{eq:betax}, we can evaluate the inverse temperature as
\begin{equation}
    \tilde{\beta}(x)=\frac{x(x+2x_1)-x_0(x_0+2x_1)}{\sqrt{x_0(x_0+2x_1)\left(1+2x_1-x_0(x_0+2x_1)\right)}}.%=\frac{(x-x_0)(x+x_0+2x_1)}{2(x_0+x_1)v_F(x_0)}.
\end{equation}
Noting that the denominator can be rewritten as $2(x_0+x_1)v_F(x_0)$, we arrive at the expression \eqref{eq:tildebetaHahnsym} reported in the main text.

In order to treat the general case, let us first define the function
\begin{equation}
    f(x) = (2x+x_1+x_2)^2v_F^2(x),
\end{equation}
which turns out to be a fourth-order polynomial in $x$. Using the coefficients \eqref{eq:ACHahn} and taking the continuum limit of the couplings, one finds explicitly
\begin{align}
\label{eq:fx}
f(x) &= 2\rho \left(x^2 (2-x_1+x_2)-x (x_1-x_2-2) (x_1+x_2)+x_1 (x_1+x_2)\right) \nonumber \\
&- \rho^2 (2 x+x_1+x_2)^2 - (x_1-x (x+x_1+x_2))^2.
\end{align}
It is easy to show that $f(x)$ has a single maximum in $x\in[0,1]$, with its location given by
\begin{equation}
    x_* = -\frac{x_1+x_2}{2} + \sqrt{(x_1+x_2)^2+8\rho\Big(1-\rho+\frac{x_2-x_1}{2}\Big)},
\end{equation}
and the value of the maximum is
\begin{equation}
    f(x_*) = R^4= 4\rho(1-\rho)(\rho+x_1)(1-\rho+x_2).
\end{equation}
We now take the derivative of the condition \eqref{eq:sinx} that has to be satisfied by $v_F(x)$ and arrive at
\begin{equation}
\frac{\pi}{\tilde L}\cos(\frac{\pi \tilde x(x)}{\tilde L})\frac{2x+x_1+x_2}{\sqrt{f(x)}}=\frac{\dd}{\dd x}\sqrt{\frac{f(x)}{f(x_*)}},
\end{equation}
where we used $\tilde x'(x)=1/v_F(x)$. Rewriting the cosine and carrying out the derivative on the r.h.s. we obtain
\begin{equation}
    \frac{\pi}{\tilde L} (2x+x_1+x_2)= \pm \frac{f'(x)}{2\sqrt{f(x_*)-f(x)}},
\end{equation}
where the $\pm$ sign applies for $x<x_*$ and $x>x_*$, respectively. This differential equation can be integrated and yields
\begin{equation}
\label{eq:fxcond}
    \sqrt{f(x_*)-f(x)} = \mp\frac{\pi}{\tilde L}\frac{1}{4}[(2x+x_1+x_2)^2-(2x_*+x_1+x_2)^2].
\end{equation}
It is a simple exercise to show, that the function in \eqref{eq:fx} indeed satisfies the above equation with the choice $\tilde L =\pi$. Finally, one can also show that \eqref{eq:fxcond} implies
\begin{equation}
    \cos\left(  \frac{\pi \tilde{x}_0}{\tilde{L}}\right)-\cos\left( \frac{\pi \tilde{x}(x)}{\tilde{L}}\right)= \frac{\sqrt{f(x_*)-f(x)}-\sqrt{f(x_*)-f(x_0)}}{R^2}= \frac{(x-x_0)(x+x_0+x_1+x_2)}{R^2},
\end{equation}
such that one indeed finds for the inverse temperature
\begin{equation}
    \tilde{\beta}(x)=\frac{(x-x_0)(x+x_0+x_1+x_2)}{(2x_0+x_1+x_2)v_F(x_0)}.
\end{equation}
\section{Functions related to the Racah chain}\label{app:racah}
The Fermi velocity associated to the Racah chain is given by
\begin{equation}
    v_F(x) = \sqrt{\frac{g_0(x) + g_1(x) \mu_0 + g_2(x) \mu_0^2}{(2 x+x_1-1)^2}}
\end{equation}
where
\begin{equation}
    \mu_0 = \rho(\rho+x_2+x_3),
\end{equation}
and the functions $g_0(x)$, $g_1(x)$ and $g_2(x)$ are given by
\begin{equation}
\begin{split}
    g_0(x) &= -\left(x^2 (x_2+x_3)+x (x_1-1) (x_2+x_3)-x_2 (x_1+x_3)\right)^2,\\
     g_1(x) &= 4 x^4+8 x^3 (x_1-1)+2x^2 \left(2 x_1^2+x_1 (x_2-x_3-6)+2 x_2 x_3-x_2-x_3+2\right)\\
     &+2x (x_1-1) (x_1 (x_2-x_3-2)+4 x_2 x_3-x_2-x_3)-(x_1-1) x_2 (x_1+x_3),\\
    g_2(x) &= -(2 x+x_1-1)^2.
    \end{split}
\end{equation}
The Fermi velocity can be rewritten as
\begin{equation}\label{eq:vf_from_beta}
     v_F(x)  = \frac{ \sqrt{1 - \Omega^2 \tilde{\beta}(x)^2 + 2 \Phi  \Omega \tilde{\beta}(x)} }{\partial_x \tilde{\beta}(x) },
    \end{equation}
where $\tilde{\beta}(x)$, $\Omega$ and $\Phi$ are given by
\begin{equation}
\begin{split}
    \tilde{\beta}(x) &=\sigma\frac{x^2+x (x_1-1)-x_0^2-x_0 (x_1-1)}{\sqrt{g_0(x_0) + g_1(x_0)\mu_0 + g_2(x_0)\mu_0^2  }},\\
   \Omega^2& = {4 \mu_0+x_2^2+2 x_2 x_3+x_3^2},\\
   \Omega \Phi & = \sigma \frac{t_0(x_0) + t_1(x_0) \mu_0 + 2\mu_0^2}{\sqrt{g_0(x_0) + g_1(x_0) \mu_0 + g_2(x_0) \mu_0^2}},
 \end{split}
\end{equation}
with $ \sigma =  \text{sign}(2 x_0 + x_1 - 1)$ and
\begin{equation}
\begin{split}
  t_0(x_0) &= (x_2+x_3) \left(x_0^2 (x_2+x_3)+x_0 (x_1-1) (x_2+x_3)-x_2 (x_1+x_3)\right) , \\
    t_1(x_0) &= \left(4 x_0^2+4 x_0 (x_1-1)+x_1 (x_2-x_3-2)+2 x_2 x_3-x_2-x_3\right).
    \end{split}
\end{equation}

\providecommand{\href}[2]{#2}\begingroup\raggedright\endgroup

\end{document}